# A Low-power Reversible Alkali Atom Source

Songbai Kang[1], Russell P. Mott[2], Kevin A. Gilmore[1,3], Logan D. Sorenson[2], Matthew T. Rakher[2], Elizabeth A. Donley[1], John Kitching[1], and Christopher S. Roper[2,a)]

1 NIST; 325 Broadway, Boulder, CO 80305; USA
2 HRL Laboratories, LLC; Malibu, CA 90265; USA
3 Department of Physics; University of Colorado; Boulder, Colorado, 80309; USA

a) author to whom correspondence should be addressed. Electronic mail: csroper@hrl.com
b) Products or companies named here are cited only in the interest of complete scientific description, and neither constitute nor imply endorsement by NIST or by the US government. Other products may be found to serve just as well.

An electrically-controllable, solid-state, reversible device for sourcing and sinking alkali vapor is presented. When placed inside an alkali vapor cell, both an increase and decrease of the rubidium vapor density by a factor of two are demonstrated through laser absorption spectroscopy on 10 to 15 s time scales. The device requires low voltage (5 V), low power (<3.4 mW peak power), and low energy (<10.7 mJ per 10 s pulse). The absence of oxygen emission during operation is shown through residual gas analysis, indicating Rb is not lost through chemical reaction but rather by ion transport through the designed channel. This device is of interest for atomic physics experiments and, in particular, for portable cold-atom systems where dynamic control of alkali vapor density can enable advances in science and technology.

Systems based on precision atomic spectroscopy currently enable some of the most precise measurements of time, frequency, rotation, and acceleration.[1,2] In many of these systems, the use of atoms cooled to microkelvin temperatures by lasers results in considerably improved performance, especially in terms of accuracy. Almost all cold-atom systems must be operated in a controlled laboratory environment, in part because the density of the alkali vapor from which the cold atoms are captured must be kept below about $4\times10^9$ cm$^{-3}$. Vapor cells containing room-temperature atoms have been scaled-down[3] for miniature traditional atomic clock implementations[4] and applications outside the laboratory. Similar work to miniaturize cold-atom systems is ongoing[5-7]. Controlling the alkali vapor density can be a significant challenge for cold-atom system applications outside the laboratory, because the density can be disturbed by long-term effects (e.g. gradual loss of alkali atoms to pumps or getters) as well as short term effects (e.g. changing temperature affecting equilibrium alkali vapor pressure). Capturing, cooling, and sustaining a population of cold atoms is only possible at alkali densities below $10^9$ cm$^{-3}$. Furthermore, the loading time of any cold atom trap is strongly dependent on the alkali density[8]. A significant step toward overcoming these challenges would be the ability to control the alkali density inside a vapor cell over relevant time scales with a reversible, electronically-actuated alkali source.

The density of a single-component vapor in thermal equilibrium with a macroscopic amount of solid or liquid is approximately an exponential function of inverse temperature within relevant temperature ranges. Alkali vapor pressure therefore can be increased or decreased by adjusting the cell temperature and, in equilibrium, is determined by the coldest point on the vapor cell walls. Using this to control alkali vapor density has the disadvantage of requiring active heating or cooling of the cell, which requires considerable power and is usually quite slow. Departures from the solid-vapor or liquid-vapor equilibrium vapor pressure can be obtained in several ways. For example, the cell can be irreversibly pumped, either actively (e.g. a vacuum pump) or passively (e.g. graphite), reducing the vapor pressure to a steady-state value balancing the pumping mechanism and evaporation from the solid or liquid source. However, irreversible pumping requires the introduction of additional atoms when high densities are required. Vapor pressures below the single-component solid-vapor or liquid-vapor equilibrium values can also be achieved through multicomponent alloys and intermetallics[9] or by limiting the alkali wall coverage to submonolayer or few monolayer equivalent. The density can be increased using a spatially separated heated region (e.g. an alkali dispenser[10,11] or an atomic beam[12]) or through the localized production of alkali atoms using a chemical reaction[13]. Light-induced atomic desorption (LIAD) can also increase the alkali vapor pressure by a factor of ten or more[14,15] and has been used to increase the loading rate of magneto-optical traps on the time scale of 100 s[16]. Pulsed alkali dispensers[17-19] and double MOT chambers[20] have also been explored for introducing and controlling alkali atom vapor density. However, none of these methods meet the requirements of being fast, reversible, low-power, miniaturizable, and contaminant-gas free which are necessary for control of atomic vapors for portable cold atomic physics applications.

Solid ion conductors when coupled to a pair of electrodes control the flux of alkali ions in the solid state[21]. When one of the electrodes is permeable to alkali atoms, these devices can be used to increase alkali vapor density[22-24]. However, the operation temperatures, voltages, and power draws remain too high for portable applications. The lowest reported operation parameters to date for a solid ionic conductor alkali source are 120 °C and 50 V.[24] This required 35 mW to operate the device in addition to the power required to maintain the device at 120 °C. This previous work also shows alkali vapor pressure reduction under reverse bias and concludes that this demonstrates alkali absorption at high activation voltages (130 V) and with widely spaced Pt electrodes. In this work, we report a solid-state electrochemical device that employs a distinctive electrode and demonstrates bidirectional rubidium vapor changes without oxygen emission. The electrodes used here support the transport and interaction of the electrons, ions, and neutrals necessary for the electrochemical oxidation of rubidium atoms across a large fraction of the device area. This enables low voltages, and when combined with a "burn-in" period to remove water, prevents the evolution of oxygen, which we have observed can cause artificial reduction in the alkali density through chemical reaction.

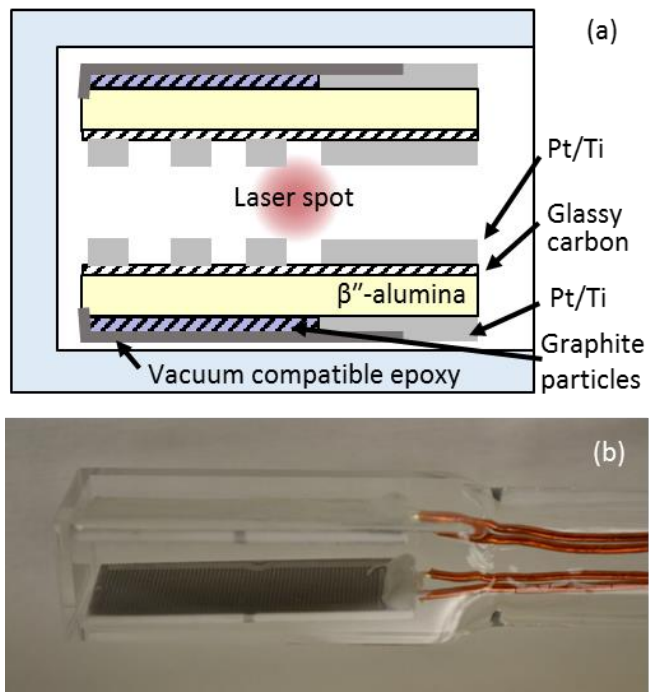

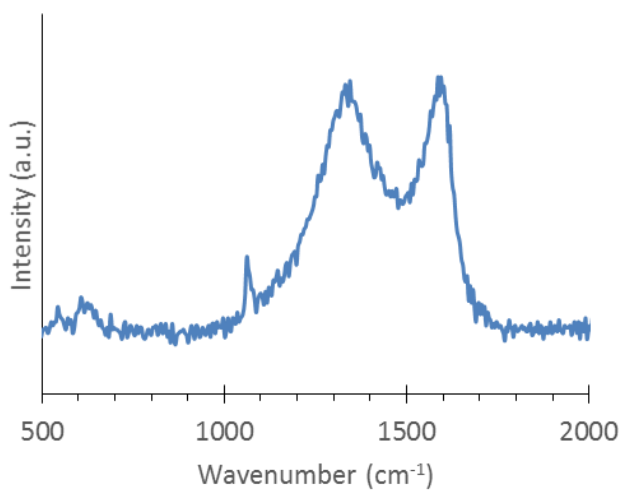

Figure 1: Solid-state electrochemical reversible alkali atom source and sink: (a) schematic of two devices in a vapor cell and (b) image of two devices in a vapor cell.

Figure 2: Raman spectrum of glassy carbon electrode on beta-alumina.

Each electrochemical device consists of a solid ion conductor sandwiched by two electrodes (Figure 1). The ion conductor is 1 mm thick sodium-beta"-alumina. Sodium-beta"-alumina was chosen instead of Rb-beta"-alumina for ease of acquisition; however, these devices are expected to interact with rubidium vapor because many metal ions, including rubidium,[25] are mobile in beta-alumina. The sodium present in the beta"-alumina is not expected to affect the transport of Rb atoms. The low vapor pressure of sodium and the selectivity of laser absorption spectroscopy should permit the observation of Rb sourcing as long as Rb sourcing experiments are preceded by a Rb sinking step to load some Rb into the device. The electrode facing the vapor consists of two layers. The layer closest to the beta"-alumina is ~10 nm thick glassy carbon. The glassy carbon material is chosen because alkali atoms can intercalate and diffuse within it. Also, graphitic carbon materials support electron conduction[26]. The glassy carbon layer thickness is chosen as a balance between minimizing electrical sheet resistance and minimizing the alkali atom capacity when it is fully saturated with alkali atoms. Reducing the alkali atom capacity is anticipated to be crucial to minimizing the time constants for sourcing and sinking alkali atoms from the vapor because diffusion through the glassy carbon cannot be started or stopped as quickly as ionic conduction. Raman spectroscopy (Figure 2) is used to confirm the disordered sp2 nature of the vapor-facing electrode through the presence of broad peaks at 1345 and 1590 cm$^{-1}$ [27]. On top of this layer are 0.1 mm wide fingers of 10 nm Ti/80 nm Pt. The backside electrode consists of 0.1 mm wide fingers of 10 nm Ti/80 nm Pt onto which graphite particles in a polymeric binder are applied. The backside electrode is encapsulated in vacuum-compatible epoxy to prevent direct interaction between alkali atoms in the backside electrode and the vapor.

The vapor cell includes two devices, each 3 cm x 0.95 cm, inside a rectangular glass cell (3.5 cm x 1 cm x 1 cm internal volume). The two devices are positioned with their vapor-facing electrodes facing each other and are separated by a gap. Four polyimide-coated copper wires connect each electrode to a unique pin on an electrical feedthrough. The open end of the glass cell is connected to a vacuum system. There is a valve for sealing off the cell containing the devices from the rest of the vacuum system, which includes a SAES[b] rubidium alkali dispenser, a turbo pump (80 L/s), and a Residual Gas Analyzer (SRS RGA 300). A 795 nm laser beam (New Focus TLB 7000) is passed through the cell between the devices. Laser frequency modulation (FM) spectroscopy ($f_{mod}$ = 10 kHz) is employed to quantify the Rb vapor density between the devices as shown in Figure 1a with its 2f component and lock laser frequency with its 1f component. After achieving a base pressure of 10$^{-7}$ Torr, electrical current is passed through the SAES Rb source emitting Rb into the system. After several hours, the Rb vapor density is ~10$^{10}$ cm$^{-3}$.

The electrochemical devices are wired in parallel. An electrical open is maintained when the devices are in an "off" state and voltage control is used when the devices are in an "on" state. Positive voltage is defined when a higher electrical potential is applied to the vapor-facing electrodes compared to the backside electrodes. Current is measured with a voltmeter across a 4.6 kΩ resistor in series with the pair of electrochemical devices.

To test for Rb sinking functionality, three square wave pulses (+20 V, 10 s duration, >230 s spacing between pulses) are applied to the device. Finally, a much longer pulse (+20 V, 400 s) is applied. Throughout this entire Rb sinking test, the SAES Rb source is continuously active and the Rb vapor density, as determined by laser absorption, is continuously recorded. The results are time shifted so that the beginning of each pulse aligns with t = 0 s and are depicted in Figure 3. For each of the 10 s pulses, a rapid drop in the Rb vapor density for approximately 13 to15 s is observed, followed by a slow rise. For the longer pulse, the Rb vapor density drops quickly in the first 15 s and continues to slowly drop as time proceeds. In all cases, current initially peaks and then decays as long as voltage is applied to the device (Figure 3b). Throughout the experiment the RGA O$_2$ signal is monitored (Figure 3c) and no significant oxygen evolution is observed.

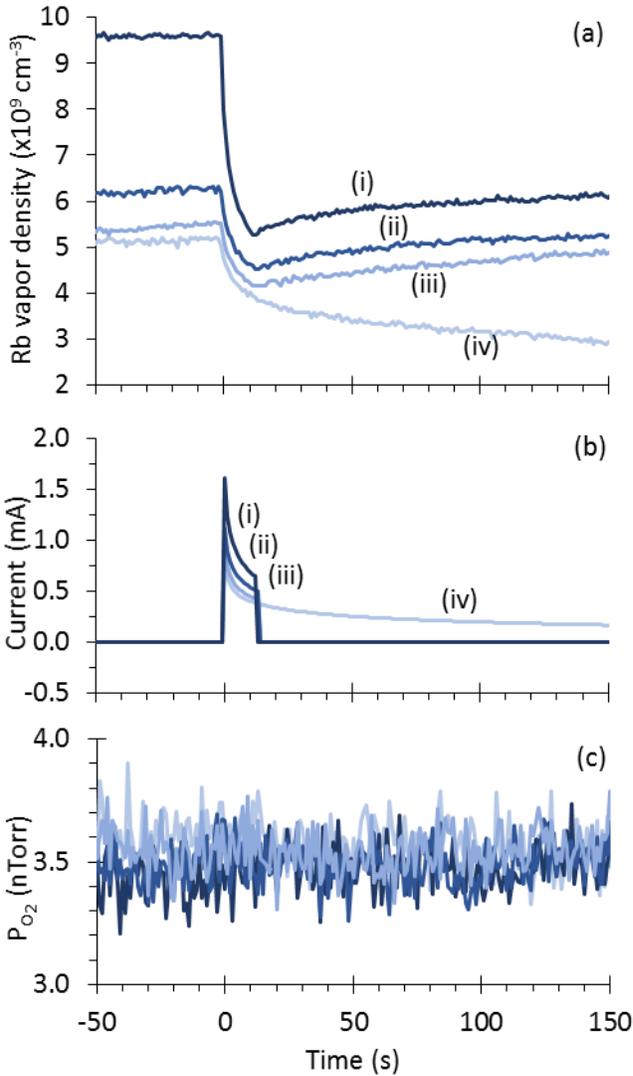

Figure 3: Testing of Rb vapor sinking capabilities of Na-beta"-alumina electrochemical device with SAES Rb source active: (a) Rb vapor density over time for (i, ii, and iii) +20 V, 10 s pulses and (iv) +20 V, 400 s pulse; (b) corresponding current vs time; and (c) corresponding RGA partial pressure of $O_2$ versus time.
5

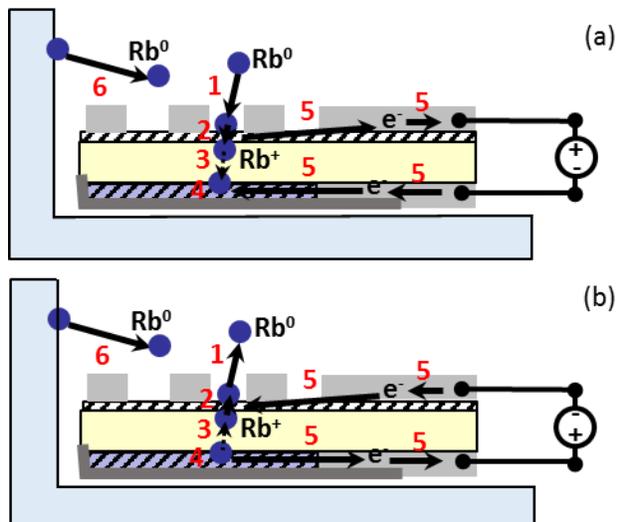

Figure 4: Mechanisms during (a) Rb vapor sinking: (1) Rb adsorption, (2) Rb diffusion in graphitic carbon, (3) $Rb^+$ ion conduction in beta-alumina, (4) Rb storage in solid graphite reservoir, (5) Electron transport in graphitic carbon and TiPt, and (6) Rb desorption from passive walls and (b) Rb vapor sourcing from walls and electrochemical device: (1) Rb desorption, (2) Rb diffusion in graphitic carbon, (3) $Rb^+$ ion conduction in beta-alumina, (4) Rb extraction from solid graphite reservoir, (5) Electron transport in graphitic carbon and TiPt, and (6) Rb desorption from passive walls. Note, when Rb vapor sourcing is from walls and not electrochemical device, only step 6 is active.

It is clear that stimulating the device causes rapid and reproducible reduction in the Rb vapor density. Because no $O_2$ emission is observed, this reduction in Rb vapor density is attributed to adsorption, electrochemical reaction, and sequestration of Rb atoms into the device rather than reaction with emitted oxygen (Figure 4a). As soon as the voltage is removed, the Rb vapor density increases. This is attributed to Rb emission from the SAES source and from Rb desorption from the glass cell walls entering the vapor region between the devices (Figure 4a). Higher magnitude initial Rb vapor densities lead to larger changes in Rb vapor density, also consistent with an adsorption process.

To test for both Rb sinking and sourcing functionality in rapid succession, additional testing is conducted with the SAES source off. Initially, the cell is loaded with Rb from the SAES source. After the Rb vapor density in the cell reaches 8 x$10^9$ $cm^{-3}$ the SAES source is turned off. Two electrical signals, each followed by a quiescent period, are applied to the device to enable comparison. First, a square pulse of +5 V for 10 s is applied followed by an open circuit across the electrochemical devices. Second, a two square pulses in immediate succession (+5 V for 10 s, then -5 V for 10 s) are applied followed by an open circuit condition. The results are time shifted so that the beginning of each positive voltage pulse aligns with t = 0 s and depicted in Figure 5. As in the prior test, the RGA $O_2$ signal is monitored (Figure 5c) and no significant oxygen evolution is observed. Of note is that only 5 V is required to activate the device. The low voltage and low current (Figure 5b) result in low peak power (<3.4 mW peak power) and low energy (<10.7 mJ per 10 s pulse).

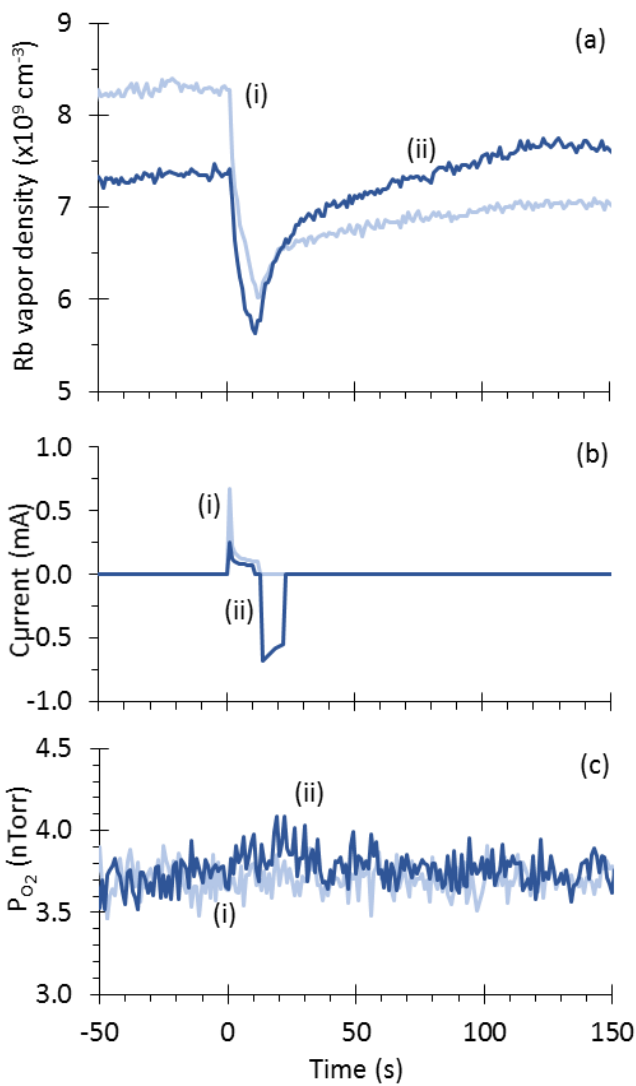

Figure 5: Testing of Rb vapor sinking capabilities of Na-beta"-alumina electrochemical device with SAES Rb source off: (a) Rb vapor density over time for (i) +5 V, 10 s pulse and (ii) +5 V, 10 s pulse followed by -5 V, 10 s pulse; (b) corresponding current vs time; and (c) corresponding RGA partial pressure of $O_2$ versus time.

Again, the application of a positive voltage reduces the Rb vapor density over 10 to 15 s and this is attributed to Rb adsorption, electrochemical reaction, and sequestration. This decrease is similar in rate and magnitude in both cases. In contrast, the increase in Rb vapor pressure that starts at around t = 15 s is faster and achieves a greater vapor density after the negative voltage pulse is applied. In the case with only the positive voltage pulse, the Rb vapor density increase is attributed to desorption of Rb from the glass cell walls. In the negative voltage pulse case, ionic transport in the electrochemical device is reversed so that alkali ions are conducted toward the vapor-facing electrode instead of being pulled toward the back electrode (Figure 4b). At the vapor facing electrode, $Rb^+$ ions can combine with electrons to form Rb neutrals. These Rb neutrals can then diffuse through the glassy carbon layer and desorb into the vapor. The Rb flux from this pathway in addition to the Rb desorption from the walls results in a greater Rb vapor density change when the negative voltage is applied.

The device test results demonstrate reversible function that depends on the polarity of the applied voltage. When the vapor facing electrode has a higher potential applied to it, the Rb vapor pressure decreases, and when the backside electrode has a higher potential applied to it, the Rb vapor pressure increases. This agrees with the expected ionic conduction directions in the beta"-alumina. Although not individually tested, the system level performance indicates that the backside electrode appears to be working as a solid-state alkali reservoir. The mechanism is assumed to be similar to lithium-ion graphite battery electrodes in which alkali atoms intercalate into graphite particles[28]. Furthermore, the system level performance indicates that the glassy carbon vapor-facing electrode is also working as intended. The glassy carbon allows both electron conduction and alkali neutral diffusion in the same material, thus enabling a large area fraction of the electrode that is electrochemically active. This is the likely reason why the device studied here functions at low voltage. This device has similar time constants compared to LIAD[15], which can change alkali vapor density in as few as 10 s; however, so far the maximum relative vapor density change of this device is below that achievable with LIAD, which can change alkali vapor density by over a factor of ten.

This device is anticipated to be useful in cold-atom systems and micro-systems. In particular, the ability to rapidly and electrically vary the alkali atom vapor density using milliwatts of power may enable control of alkali atom vapor density in demanding thermal environments and over long durations. This can be especially useful for portable cold-atom systems. Significantly faster device operation is likely possible with optimized electrode chemistries and designs. Because many alkali metals, alkaline earth metals, as well as some transition metals and metalloids form mobile ions in beta-alumina, this device is useful for more than rubidium vapor. Furthermore, the ability to rapidly vary alkali vapor density may enable new kinds of cold atom trapping and measurement sequences.

This material is based upon work supported by the Defense Advanced Research Projects Agency (DARPA) and Space and Naval Warfare Systems Center Pacific (SSC Pacific) under Contract No. N66001-15-C-4027. The authors acknowledge DARPA program manager Robert Lutwak as well as Jason Graetz, John Vajo, Adam Gross, Rick Joyce, Danny Kim, Randall Schubert, and Brian Cline of HRL Laboratories, LLC for useful discussions. We further acknowledge Florian Herrault, Tracy Boden, Margie Cline, Ryan Freeman, and Lian-Xin Coco Huang for assistance with device fabrication as well as Kay Son for assistance with Raman spectroscopy. This work is a contribution of NIST, an agency of the U.S. Government, and is not subject to copyright.

1    Ch. J. Bordé, Metrologia **39** (5), 435 (2002).
2    J. Kitching, S. Knappe, and E. A. Donley, IEEE Sensors J **11** (9), 1749 (2011).
3    L.-A. Liew, S. Knappe, J. Moreland, H. Robinson, L. Hollberg, and J. Kitching, Appl Phys Lett **84** (14), 2694 (2004).
4    S. Knappe, V. Shah, P. D. D. Schwindt, L. Hollberg, J. Kitching, L.-A. Liew, and J. Moreland, Appl Phys Lett **85** (9), 1460 (2004).
5    J. A. Rushton, M. Aldous, and M. D. Himsworth, Rev Sci Instrum **85** (12), 121501 (2014).
6    V. Shah, R. Lutwak, R. Stoner, and M. Mescher, Proceedings of the 2012 IEEE International Frequency Control Symposium, 2012, pp. 1-6.
7    J. Sebby-Strabley, K. Salit, K. Nelson, J. Ridley, and J. Kriz, Proceedings of the 43rd Annual Precise Time and Time Interval Systems and Applications Meeting, Long Beach, California, 2011, pp. 231-238.
8    M. Prentiss, E. L. Raab, D. E. Pritchard, A. Cable, J. E. Bjorkholm, and S. Chu, Opt Lett **13** (6), 452 (1988).

9       M. B. Squires, Ph.D. thesis, University of Colorado, 2008.
10      C. Wieman, G. Flowers, and S. Gilbert, Am J Phys **63** (4), 317 (1995).
11      M. Succi, R. Canino, and B. Ferrario, Vacuum **35** (12), 579 (1985).
12      W. D. Phillips and H. Metcalf, Phys Rev Lett **48** (9), 596 (1982).
13      W. Espe, *Materials of High Vacuum Technology*. (Pergamon, Oxford, 1966).
14      M. Meucci, E. Mariotti, P. Bicchi, C. Marinelli, and L. Moi, Europhys Lett **25** (9), 639 (1994).
15      A. Burchianti, A. Bogi, C. Marinelli, E. Mariotti, and L. Moi, Physica Scripta **2009** (T135), 014012 (2009).
16      B. P. Anderson and M. A. Kasevich, Phys Rev A **63** (2), 023404 (2001).
17      J. Fortagh, A. Grossmann, T. W. Hänsch, and C. Zimmermann, J Appl Phys **84** (12), 6499 (1998).
18      U. D. Rapol, A. Wasan, and V. Natarajan, Phys Rev A **64** (2), 023402 (2001).
19      V. Dugrain, P. Rosenbusch, and J. Reichel, Rev Sci Instrum **85** (8), 083112 (2014).
20      C. J. Myatt, N. R. Newbury, R. W. Ghrist, S. Loutzenhiser, and C. E. Wieman, Opt Lett **21** (4), 290 (1996).
21      P. P. Kumar and S. Yashonath, J Chem Sci **118** (1), 135 (2006).
22      F. Gong, Y.-Y. Jau, K. Jensen, and W. Happer, Rev Sci Instrum **77** (7), 076101 (2006).
23      J. J. Bernstein, S. Feller, A. Ramm, J. North, J. Maldonis, M. Mescher, W. Robbins, R. Stoner, and B. Timmons, Solid State Ionics **198** (1), 47 (2011).
24      J. J. Bernstein, A. Whale, J. Brown, C. Johnson, E. Cook, L. Calvez, X. Zhang, and S. W. Martin, Proceedings of the Solid-State Sensors, Actuators and Microsystems Workshop, Hilton Head Island, South Carolina, 2016, pp. 180-184.
25      S. J. Allen, A. S. Cooper, F. DeRosa, J. P. Remeika, and S. K. Ulasi, Phys Rev B **17** (10), 4031 (1978).
26      T. Zheng, Y. Liu, E. W. Fuller, S. Tseng, U. von Sacken, and J. R. Dahn, J Electrochem Soc **142** (8), 2581 (1995).
27      M. I. Nathan, J. E. S. Jr., and K. N. Tu, J Appl Phys **45** (5), 2370 (1974).
28      M. Yoshio, R. J. Brodd, and A. Kozawa, *Lithium-Ion Batteries: Science and Technologies*. (Springer, New York, 2009).